\documentstyle[prb,aps]{revtex}

\input{epsf}

\begin{document}
\draft
\title{Spin quantization axis dependent magnetic properties 
and x-ray magnetic circular dichroism of 
FePt and CoPt}
\author{I.~Galanakis,  M.~Alouani, and H.~Dreyss\'e}
\address{IPCMS-GEMME, 23, rue du Loess, F-67037
Strasbourg Cedex, France}
\maketitle

\begin{abstract}
We have performed  a theoretical study of the  magnetic
circular dichroism in the x-ray absorption spectra (XMCD) of the
equiatomic CoPt and FePt ordered alloys as a function of
the spin quantization axis. We found that the magnetization axis is
along the [001] direction 
and the magneto-crystalline anisotropy energy (MCA) for the FePt compound
is twice as large as that of the CoPt compound in agreement with 
experiment. The band structure and the total
density of states confirm that all electronic states contribute to
the MCA, and not just the states at the vicinity of the Fermi level.
The orbital magnetic moments decrease with respect to the angle
between the [001] axis and the spin quantization axis, and are
much larger 
for the CoPt compound. We show that the orbital moment anisotropy 
is reflected in the XMCD signal.
\end{abstract}


\section{Introduction}

The disordered equiatomic binary alloys of the XY (X= Fe, Co -- Y= Pd, Pt) type 
crystallize in the fcc structure 
and the magnetization  is along the [111]
axis.\cite{razee}  
At low temperatures these alloys 
tend to order in the $L1_0$ layered-ordered
structure and in this case  the spontaneous magnetization tends to align
 perpendicular to the layer stacking explaining the behavior
of CoPt films during the magnetic annealing.\cite{razee2} 
The strong perpendicular magnetic anisotropy (PMA) is due to the
highly anisotropic $L1_0$ structure and it   makes them very
attractive for magnetic recording devices.\cite{coffey}
 The latter    structure can be also obtained by
molecular beam epitaxy (MBE) of alternating layers of pure X and Y atoms
due to the substrate induced constraints. The first observation of
the $L1_{0}$ long-range order for a [001] CoPt film grown by MBE
was made by Harp {\em et al.}\cite{harp} in 1993 and for a [001]
FePt film by Cebollada {\em et al.}\cite{cebollada} Lately other techniques 
have been also employed to develop films presenting 
PMA. 
 CoPt  films were grown by sputtering by Visokay {\it et
al.}\cite{visokay} and by evaporation by Lin and Gorman,\cite{lin} and
FePt films were grown by  various sputtering
techniques.\cite{watanabe,lairson,watanabe2,mitani,sato}

The magneto-crystalline anisotropy energy (MCA)  can be probed
by many techniques such as torque or ferromagnetic resonance
measurements. Both these methods describe
the MCA in terms of phenomenological
 anisotropy constants. It has been demonstrated by Weller 
{\em et al.}\cite{weller} that x-ray
 magnetic circular dichroism (XMCD) is also a suitable technique 
for probing the MCA,
\textit{via} the determination of the anisotropy of the orbital
magnetic moment on a specific shell and site.  The x-ray
absorption spectroscopy (XAS) using polarized radiation
 probes element specific magnetic properties of alloys 
by applying the XMCD 
 sum rules to the experimental
spectra.\cite{thole,carra,VanL98} However for 
itinerant systems, in particular to low symmetry systems,
the use of the sum rules is debated because they are derived from an 
atomic 
theory.\cite{carra,chen,wu}
Lately angle-dependent XMCD experiments have been used to provide 
a deeper understanding for the relation between MCA and the orbital 
magnetic moments.\cite{grange2}

The x-ray absorption for CoPt multi-layers has been already
studied experimentally by Nakajima {\em et al.},\cite{naka} Koide
{\it et al.},\cite{koide} R\"uegg {\it et al.},\cite{ruegg}   and
Sch\"ultz {\it et al.}\cite{schultz} 
Nakajima {\em et
al.}\cite{naka} revealed a strong enhancement of the cobalt orbital
moment when PMA was present, Koide {\it et al.}\cite{koide}
showed that with decreasing cobalt
thickness the easy axis rotates from in-plane to out-of-plane, and
R\"uegg {\it et al.}\cite{ruegg} that platinum polarization increases also with decreasing cobalt
thickness. Hlil {\it et al.}\cite{hlil} showed by x-ray absorption
spectroscopy that modifications of platinum edges in different compounds
are correlated to the change in the number of holes. 

Several {\em ab-initio}
calculations have already been performed to investigate the
XMCD.\cite{Ebert,Meb1,Brouder} The $L_2$- and $L_3$-edges involving
electronic excitations of 2$p$-core electrons towards $d$-valence
states have primarily attracted much attention due to dependence
of the dichroic spectra on the exchange-splitting and the
spin-orbit coupling of both initial core and final valence states.
For 5$d$ elements dissolved in 3$d$ transition metals, the
spin-orbit coupling of the initial 2$p$-core states is large and
the resulting magnetic moment is small, while the opposite is true
for the 3$d$ elements. This can lead to a pronounced dichroic
spectra as seen by Sch\"utz in the case of 5$d$ elements dissolved
in iron.\cite{schultz} 

In this work we study the correlation between the quantization axis dependent
XMCD and the magnetic properties of both ordered alloys 
FePt and CoPt. Our method is based on an 
all-electron relativistic and spin-polarized full-potential
muffin-tin orbital method (LMTO)\cite{WILLS,LMTO} in conjunction
with   both the von Barth and Hedin parameterization to the local
density approximation (LSDA)\cite{barth}  and the generalized
gradient approximation (GGA)\cite{perdew} to the exchange
correlation potential. 
The implementation of the calculation of the XMCD spectra 
has been presented in a previous work.\cite{Meb1}
In section 2 we present the details of the 
calculations, while in sections 3
and 4  we discuss  our MCA and the magnetic spin and orbital
moments, respectively. In section 5  we present 
our calculated  XMCD as a function of the spin quantization axis, 
and in section 6 we discuss  the interpretation of the MCA using
the band structure and the total density of states anisotropy.

\section{Computational Details}

  To compute the electronic properties of CoPt and FePt we used
the experimental lattice constants ($a$=3.806\AA \ and
$c$/$a$=0.968 for CoPt\cite{grange2}; $a$= 3.861\AA \ and
$c$/$a$=0.981 for FePt\cite{villars}) and a unit cell containing
one atom of cobalt(iron) and one of platinum. The $L1_0$ structure  can be
seen as a system of alternating  cobalt(iron) and platinum layers along the
[001] direction.  The MCA and the XMCD are computed with respect to 
the angle $\gamma$ between the [001] axis and the spin
quantization axis on the (010) plane. So $\gamma$=0$^o$
corresponds to the [001] axis and $\gamma$=90$^o$ to the [100] axis.

MCA can be computed directly using  {\em ab-initio} methods; it is
defined as the difference between the total energy for two
different spin quantization axis. The spin-orbit coupling contribution
to MCA is implicitly included in our {\em ab-initio} calculations,
and we do  not take into account the many-body
interactions of the spin-magnetic moments\cite{janssen} since their
contribution to the MCA is negligible.\cite{bruno2} The number of {\bf
k} points for performing the  Brillouin zone (BZ) integration
depends strongly on the interplay between the contributions to the MCA
from the Fermi surface and the remaining band structure contribution to the total
energy.\cite{Sol} When the former contribution to the MCA is
important, a large number of {\bf k}-points is needed to describe
accurately the Fermi surface. For the two studied systems we found
that 6750 {\bf k}-points in the BZ are enough to converge the MCA within 0.1 meV.

To perform the integrals over the BZ we use a Gaussian broadening
method which convolutes each discrete eigenvalue with a Gaussian
function of width 0.1 eV. This method is known to lead to a fast
and stable convergence of the spin and charge densities compared
to the standard tetrahedron method. To develop the potential
inside the MT spheres we calculated a
 basis set of lattice harmonics including functions up  to
 $\ell=8$, while for the FFT we
 used a real space  grid of 16$\times$16$\times$20.
 We used a double set of basis functions, one set to describe the valence
 states and one for the unoccupied states.
 For the valence electrons we used a basis set containing
3$\times s$, 3$\times p$ and 2$\times d$ wave functions, and for
the unoccupied states 2$\times s$, 2$\times p$ and 2$\times d$
wave functions.

\section{Magneto-crystalline Anisotropy}

\begin{figure}
\begin{minipage}{6.0in}
\epsfxsize=5.0in \epsfysize=4.0in \centerline{\epsfbox{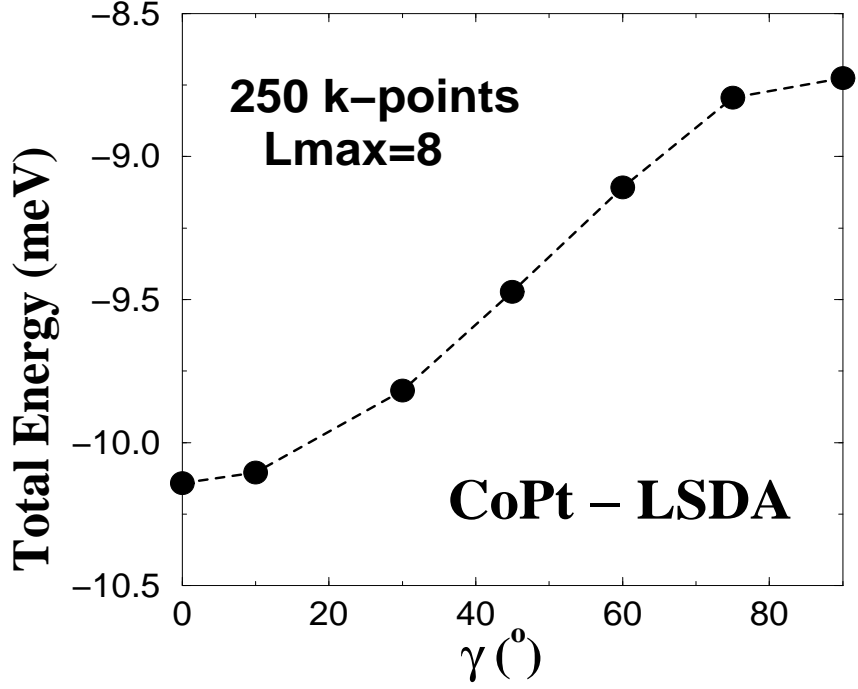}}
\end{minipage}
\caption{ \label{fig12}  Total energy for the CoPt system within
LSDA  as a function of the angle $\gamma$ between the [001] axis
and the spin quantization axis on the (010) plane. The fundamental
state corresponds to $\gamma$=0$^o$, that means to the [001] axis.
The total energy increases as a function of $\gamma$.}
\end{figure}

\begin{figure}
\begin{minipage}{6.0in}
\epsfxsize=5.0in \epsfysize=4.0in \centerline{\epsfbox{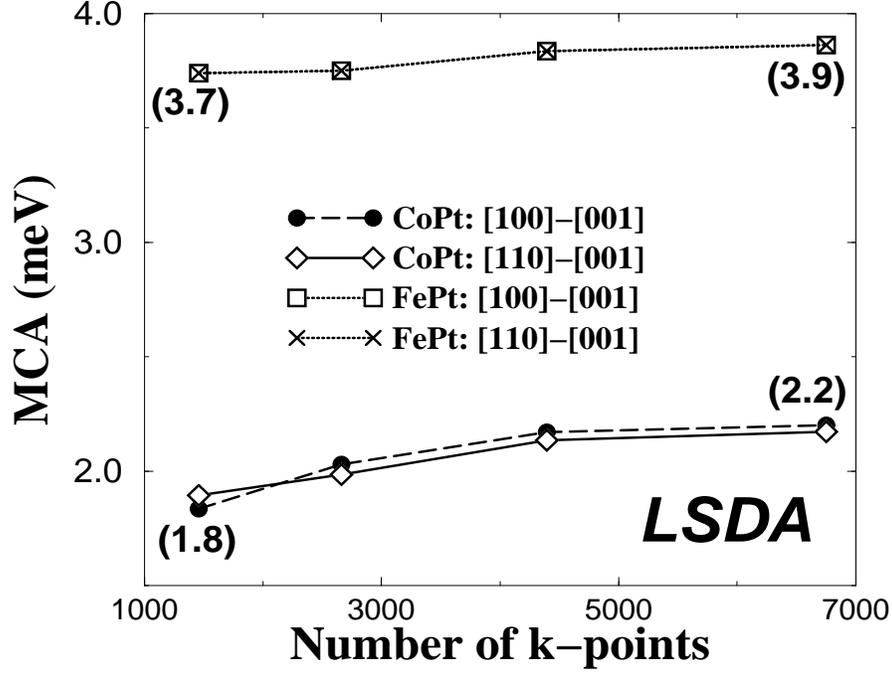}}
\end{minipage}
\caption{ \label{fig2}
Convergence of the magneto crystalline anisotropy in meV with 
respect to the number of
{\bf k}-points. The  calculation is performed for two in-plane axis. For CoPt
the line for the [100] axis converges to 2.2 meV while the line for
the [110] axis converges to 2.17 meV.  For FePt the two lines differ by 
less than 0.01 meV. The behavior of
MCA verifies the assumption that the $L1_0$ structure is isotropic
in the (001) plane.}
\end{figure}

CoPt and FePt films are known to present a strong uniaxial MCA,
because of the high anisotropic $L1_0$ inter-metallic phase. 
Experimentally the magnetization axis is found to be along the [001]
direction.\cite{eurin,farrow} In a first step we performed
calculations with 250 {\bf k}-points in the BZ. This
number of {\bf k}-points  is large enough to produce accurate total
energy when  Gaussian smearing is used for the
integration in the Brillouin Zone but not enough accurate to
compute the MCA. In figure\ \ref{fig12} we present
calculations with respect to the angle $\gamma$ between the [001]
axis and the spin quantization axis on the (010) plane for the
CoPt compound within the LSDA. We observe that total energy value increases 
with the angle $\gamma$. So the ground state corresponds to $\gamma$=0$^o$,
{\em i.e.} the [001] axis. The same behavior occurs for  the FePt
compound. Because  250
{\bf k}-points are not enough to produce  an accurate value for the
MCA, we present in figure \ref{fig2} the convergence of the MCA with the number of {\bf
k}-points  for both
compounds within LSDA. The MCA is the difference in the total
energy between the in-plane axis [100] and [110] and the easy axis
[001]. We have found that the total energy difference between
the [100] and the [110] directions is negligible compared to the difference 
between the in-plane axis and the [001] axis.
 Our values are converged up to 6750 {\bf k}-points and are
given per unit cell (one atom of X and one of platinum).
 The MCA converges to 2.2 eV for CoPt and 3.9 eV for FePt.
These behavior confirms the assumption that the system is
isotropic inside the plane. The  GGA MCA calculations
converged to 1.9 meV and 4.1 meV for CoPt and FePt,
respectively. The GGA results seem to be in good agreement 
with the LSDA results.

Daalderop {\em et al.}\cite{Daal} performed calculations for MCA
in CoPt and FePt by means of an LMTO method in the atomic sphere
approximation (ASA) within LSDA\cite{LMTO}  using the 
 force theorem.\cite{macintosh} The easy magnetization axis found 
by this latter calculation is the [001] for both systems in agreement
with our LSDA and GGA results. Their  MCA value is  2 meV 
for CoPt and is 3.5 meV for FePt. 
Solovyev {\em et al.}\cite{Sol} used the same structure within a 
real-space Green's
function technique framework to find also that the magnetization
is along the [001] axis. Including  the spin-orbit
interaction for all the atoms, they found a CoPt MCA value of 2.3
meV and a FePt value of 3.4 meV. Our computed value agrees also
with the value of 1.5 meV for CoPt and 2.8 meV for FePt, obtained
by Sakuma\cite{Sak} using LMTO method in the atomic sphere
approximation (ASA) in conjunction with the force
theorem.\cite{macintosh} The drawback of the LMTO-ASA method is that it 
accounts only for the spherical part of the potential and 
ignores the interstitial region.
Furthermore, the force theorem does not  account directly for the
exchange-correlation contribution to the MCA. Finally, Oppeneer used the
augmented spherical waves method (ASW) in the atomic 
sphere approximation and found a MCA value of
2.8 meV for FePt and 1.0 meV for CoPt,\cite{oppen} which are the
smallest among all the {\em ab-initio} calculations.

 Grange {\em et al.},\cite{grange2} using a  
torque measurement for a MBE deposited CoPt film on a MgO(001)
substrate,  obtained a MAE of 1.0 meV. An early measurement of 
a monocristal of CoPt, by  Eurin and Pauleve, produced a value 
of 1.3 meV.\cite{eurin} The large 
value of Eurin and Pauleve is due to the fact that their
sample was completely ordered. For FePt the first experiment of
Ivanov {\em et al.}\cite{ivanov} produced an anisotropy value
of 1.2 meV and showed that for a thin film the shape
anisotropy would be one order of magnitude smaller compared to
the MCA. Farrow {\em et al.}\cite{farrow} and Thiele {\em et
al.}\cite{thiele} found for MBE deposited FePt films on a MgO(001)
substrate an anisotropy value of 1.8 meV. These films were highly
ordered (more than 95\% of the atoms were in the correct site).
All experiments have been carried out at room temperature,  which
explain at some extend 
the difference between the calculated and experimental MCA values
(the MCA decreases with temperature\cite{eurin}).
It is worth mentioning that the experimental MCA for FePt is 
much larger than that of CoPt in agreement with our calculations.
For thick films volume shape
anisotropy (VSA) contributes also to the MAE and it favors always an
in-plane magnetization axis. We  can estimate the VSA as
$-2\pi M_V^2$ in c.g.s. units, where $M_V$ is the mean magnetization
density,\cite{bruno2} and obtain a  value of  -0.1 meV for FePt
and -0.06 meV for CoPt. These values are one order of magnitude
smaller than the MCA values in agreement with the speculation
of  Ivanov {\em et
al.}\cite{ivanov} 

\section{Magnetic moments}
\subsection{Density of States}

\begin{figure}
\begin{minipage}{6.0in}
\epsfxsize=5.0in \epsfysize=4.0in \centerline{\epsfbox{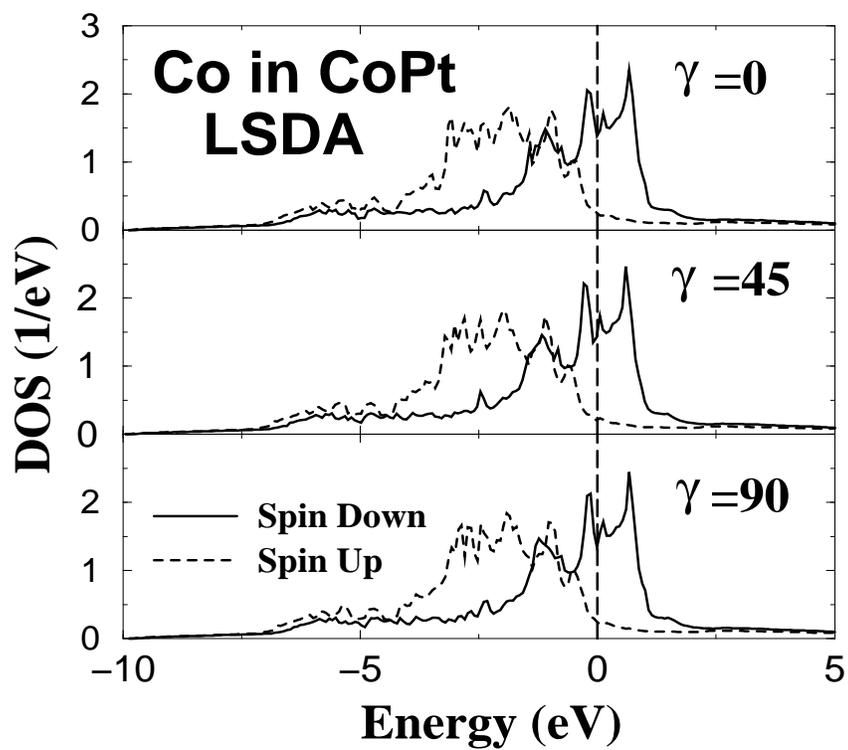}}
\end{minipage}
\caption{ \label{fig3} Cobalt projected partial density of states with
respect to the angle $\gamma$ between the [001] axis and 
the spin quantization axis in the (010) plane. There are no major changes with $\gamma$.
This behavior reflect the isotropic character of the
spin moments.
}
\end{figure}

In figure\ \ref{fig3} we present the cobalt-projected partial
density of states for three spin quantization axis corresponding
to angles $\gamma$= 0$^o$, 45$^o$ and 90$^o$ calculated with 6750
{\bf k}-points. The 3-$d$ states dominate the electronic structure
of cobalt. The spin-up band is practically totally occupied while the
spin down band is almost half-occupied. The general form of the cobalt DOS,
as well as the iron DOS, does not seem to change appreciably with the 
angle $\gamma$. 
The platinum projected density of states show similar behavior for 
both FePt and CoPt compounds. 
 Bulk platinum is paramagnetic and the small changes in the
DOS come from the polarization of the 5$d$ electrons via
hybridization with the 3$d$ electrons of cobalt (iron). It is worth
mentioning that the DOS is calculated inside each muffin-tin and the
interstitial region is not taken into account. To minimize the
contribution of the interstitial region we use almost touching 
 muffin-tin spheres. In addition, we have  find that remaining 
region has a negligible spin polarization.

\subsection{Spin Magnetic Moments}
The spin magnetic moments are isotropic with respect to the
spin quantization axis as expected. They were  calculated  by attributing
all the charge inside each muffin-tin sphere to the atom located in that
sphere.  As outlined above we have found that  the interstitial contribution 
to the spin magnetic moment is one order of
magnitude smaller  than that of the platinum site. In CoPt
we found a cobalt spin magnetic moment of 1.79 $\mu_B$ and a  platinum
moment of 0.36 $\mu_B$ within the LSDA. The GGA cobalt spin 
magnetic moment of 1.83 $\mu_B$ is slightly larger than the value 
within LSDA.  This is due to a more
atomic like description of the atoms in a solid within the GGA compared to 
the  LSDA.
Although the GGA underestimates the hybridization between cobalt and platinum
$d$ valence electrons compared to the LSDA, the larger cobalt spin moment
leads to a slightly larger GGA platinum spin moment of 0.37 $\mu_B$. Our
calculated values are in good agreement with the experimental values
of Grange {\em et al.} (1.75$\mu_B$ for cobalt and
0.35$\mu_B$ for platinum).\cite{grange2}
Previous experiments by van Laar on a
powder sample gave a value of 1.7 $\mu_B$ for the cobalt atom and 0.25
for the platinum atom.\cite{laar} The spin magnetic moments have been
previously calculated by Solovyev {\em et al.}\cite{Sol} (1.72 for cobalt and
0.37 for platinum), by Daalderop {\em et al}\cite{Daal} (1.86 for cobalt), by
Sakuma\cite{Sak} (1.91 for cobalt and 0.38 for platinum), and finally by
Kootte {\em et al.}\cite{kootte} by means of a localized spherical wave
method (1.69 for cobalt and 0.37 for platinum). All previous
calculations are in good agreement with our full-potential results.

As expected the iron spin moments are much larger than the cobalt ones.
The LSDA produced  a
value of 2.87 $\mu_B$ while the GGA produced  a slightly 
larger value, 2.96 $\mu_B$. The hybridization between the iron 3$d$
states and the platinum 5$d$ states is less intense than in the case of
CoPt resulting in a smaller platinum moment for FePt. The LSDA platinum spin moment
is 0.33 $\mu_B$ (compared to 0.36 $\mu_B$ in CoPt) and the GGA platinum
spin moment is 0.34 $\mu_B$ (compared to 0.37 $\mu_B$ in CoPt). The spin
magnetic moments have been previously calculated by Solovyev et
al\cite{Sol} (2.77 for iron and 0.35 for platinum), by Daalderop et
al\cite{Daal} (2.91 for iron), by Sakuma\cite{Sak} (2.93 for iron and
0.33 for platinum), and finally by Osterloch {\em et al.}\cite{osterloch} (2.92
for iron and 0.38 for platinum). Here again all
previous calculations are in good agreement with our
full-potential results. All the methods produced a smaller induced spin moment
for platinum atom in FePt than for CoPt, verifying that the hybridization
effect in CoPt is much stronger than in FePt.

\subsection{Orbital Magnetic Moments}

\begin{figure}
\begin{minipage}{6.0in}
\epsfxsize=5.0in \epsfysize=4.0in \centerline{\epsfbox{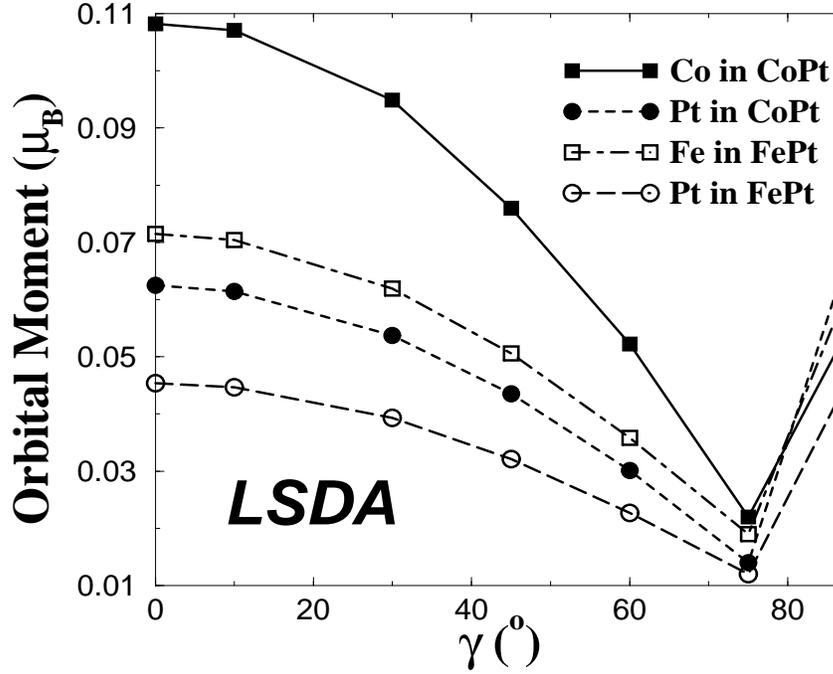}}
\end{minipage}
\caption{ \label{fig4}
Calculated orbital moments  for both compounds within LSDA as a function of the angle 
$\gamma$ between the [001] direction and the spin quantization axis 
(the lines are guides for the eye). All
four atoms show the same behavior. The orbital moments decrease with
the angle $\gamma$ until about 75$^o$, then they increase. 
The moments for the CoPt compound are
larger than for these of FePt. 
}
\end{figure}

Contrary to spin moments, the orbital moments are anisotropic. 
Figure  \ref{fig4} presents the behavior of the orbital moments as
 a function of the angle $\gamma$ between the [001] direction and 
the spin quantization axis within LSDA for both CoPt and FePt compounds 
(the lines are guide to the eye). The orbital
moments decrease with respect to the angle $\gamma$ but the values for the
[100] axis does not follow this general trend. All four lines seem to have the
same behavior, but it is interesting to notice  that for the CoPt and
magnetization along the [100] axis cobalt orbital moment is smaller
than the platinum one (for the values see Tables \ref{table1} and
\ref{table2}). The cobalt moments decrease faster than platinum moments in CoPt.
For the iron and platinum atoms in FePt the two lines are practically
parallel. The platinum orbital moments in FePt are smaller than for the CoPt
with a factor that varies from  73\% for $\gamma$=0$^o$ to 67\%
for $\gamma$=90$^o$. The ratio of iron and platinum orbital moments
in FePt varies from 1.56 for $\gamma$=0$^o$ down to 1.35 for for
$\gamma$=90$^o$, which is considerably smaller than the ratio
of cobalt and platinum orbital moments in CoPt. We notice here
that in our calculations we can estimate only the projection of
the total orbital moment on the spin quantization axis and we have
no information concerning the real value of the total magnetic
moment or its direction in space. It seems that for the magnetization
along the [100] direction, the direction of the orbital
magnetic moment undergoes a discontinuous jump, resulting in a
large projection on the spin quantization axis with respect 
to $\gamma$ at the vicinity of  90$^o$. This behavior is 
also reproduced by the GGA.

In Table \ref{table1} we present the values 
of the orbital moments of iron and cobalt within both the LSDA and
GGA as a function of the angle $\gamma$.  
The  GGA values seem to be smaller than the LSDA ones but
follow exactly the same trends. The orbital moment anisotropy is
more important in the case of cobalt. The LSDA cobalt orbital moment changes
by 0.048 $\mu_B$  and the GGA moment by 0.027 $\mu_B$ as we pass from
the easy axis [001] to the hard axis [100]. The LSDA iron moment
changes by 0.002 $\mu_B$ and the GGA moment is the same for the two
high symmetry directions. In the case of cobalt the  difference
between values calculated within the two functionals, LSDA and
GGA, becomes smaller when the angle increases and for 
75$^o$ it changes sign.
In the
case of iron this difference decreases only slightly with the angle but since the
difference is considerably smaller than in CoPt (less than
0.004), we conclude that the orbital moment in the LSDA and the GGA are
roughly the same.

In Table \ref{table2} we present the values for the platinum orbital
moment within both functionals. We see that the GGA produces larger
moments than the LSDA for platinum in FePt contrary to CoPt. The platinum moments are
in general smaller than the moments of the 3$d$ ferromagnets, 
and the difference between
the values calculated within LSDA and GGA are small.
The absolute values for platinum are comparable to 
cobalt(iron) orbital moments even though the spin moments
on platinum are one order of magnitude smaller than for cobalt(iron). The
large orbital moments for platinum are due to a much larger spin-orbit coupling
for the $d$ electrons of the platinum compared to the 3$d$ ferromagnets.

The orbital moments of FePt and CoPt have been previously calculated by Daalderop and
collaborators\cite{Daal} and by Solovyev and
collaborators\cite{Sol} for the [001] direction using the LSDA. 
The orbital moment of the cobalt site was found to be 0.12 $\mu_B$ 
by  Daalderop and  0.09 $\mu_B$ by Solovyev. The
value of Daalderop is closer to our LSDA value of 0.11
$\mu_B$. For iron site Daalderop found a value of 0.08 $\mu_B$ and Solovyev 0.07
$\mu_B$ in good agreement with our LSDA value. The  platinum orbital moment 
has been calculated by Solovyev. 
 He found a value of 0.06 $\mu_B$ for platinum in
CoPt and 0.044 $\mu_B$  for platinum in FePt, close to our values, 0.06 $\mu_B$ and
0.05 $\mu_B$, respectively.

On the other hand, experimental data are available for CoPt by Grange;\cite{grange2}
obtained by applying the sum rules to the experimental XMCD spectra. The sum rules
give the moments per hole in the $d$-band. To compare experiment with
theory we calculated   the number of $d$-holes by integrating the
$d$ projected density of states inside each muffin-tin sphere. We
found  2.63 $d$  and 2.48 $d$ holes for cobalt and 
platinum, respectively. The cobalt orbital moment varies from 0.26 $\mu_B$ 
for $\gamma$=10$^o$ down to 0.11 $\mu_B$ for $\gamma$=60$^o$. The measured values are also 
available for two other angles: 
 0.24 $\mu_B$ for
$\gamma$=30$^o$ and 0.17 $\mu_B$ for $\gamma$=45$^o$. Our theory
reproduces qualitatively the experimental trends but underestimates the absolute values 
by more than
50\% (see Tables \ref{table1} and \ref{table2} for all the values). 
The calculated values show a less sharp decrease
with the angle than the experimental ones. For platinum the experimental data are
available only for two angles 10$^o$ and 60$^o$. For 
$\gamma$=10$^o$ the orbital moment is 0.09 $\mu_B$ and for
$\gamma$=60$^o$ it is 0.06 $\mu_B$. For the platinum site 
the calculated values are in much better agreement
than for the cobalt site but we must keep in mind that the sum rules have
their origin in an atomic theory and their use for 5$d$ itinerant
electrons like in platinum is still debated.
The discrepancy between
the theory and the experiment comes mainly from the 
approximation to the exchange and correlation. Both
the LSDA and the GGA approximations  to the density functional theory are
known to underestimate the orbital moment values, because the orbital
moment is a property directly associated with the current in the
solid and a static image is not sufficient. But until now a DFT
formalism like the current and spin density functional
theory (CS-DFT),\cite{vignale} which can treat at the same footing
the Kohn-Sham and the Maxwell equations is too heavy to implement in
a full-potential {\em ab-initio method}. The other 
 problem is that there is no  form of the exchange
energy of a homogeneous electronic gas in a magnetic
field known and this is the main quantity entering the CS-DFT
formalism. Brooks has also developed an {\it ad hoc} correction to the  
Hamiltonian   to account for the
orbital polarization  but this correction originates from
an atomic theory and its application to itinerant systems is 
not satisfactory.\cite{brooks}

\section{XMCD}

XMCD spectroscopy became popular after the development of the sum
rules that enable the  extraction of   reliable information on the
micro-magnetism directly from the experimental
spectra.\cite{thole,carra} The great advantage of XMCD is that we can
probe each atom and orbital in the system  so to obtain  information 
on the local magnetic properties. Lately angle-dependent XMCD
experiments  allowed the determination 
of magnetic properties for different spin quantization axis. 
All experimental spectra for
CoPt have been obtained by Grange {\it et al},\cite{grange2} and all
calculated spectra presented in this section were obtained using the LSDA.
The GGA produced the same results and are not presented. 

\begin{figure}
\begin{minipage}{6.0in}
\epsfxsize=5.0in \epsfysize=4.0in \centerline{\epsfbox{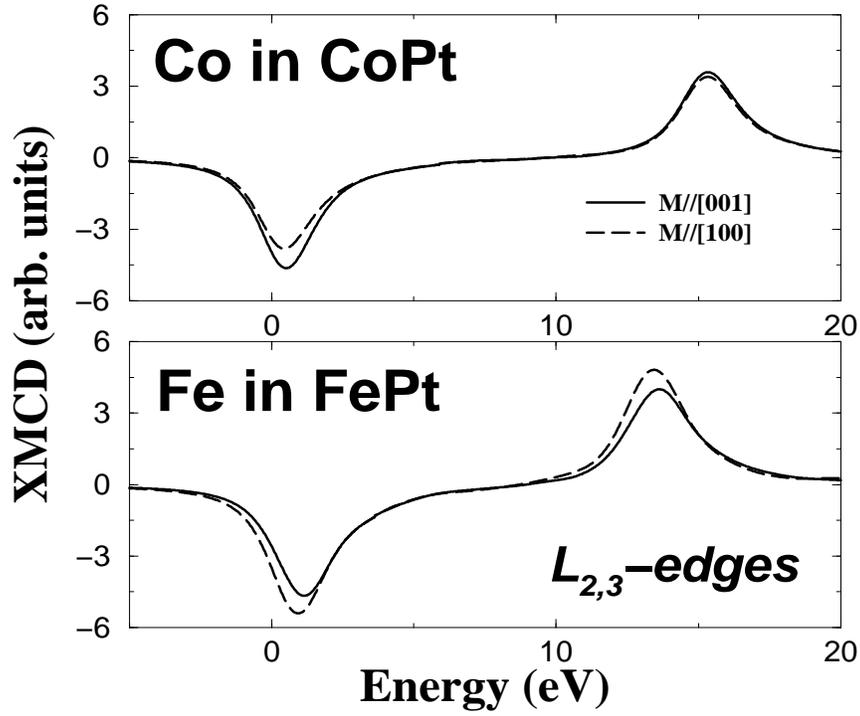}}
\end{minipage}
\caption{ \label{fig5}
 XMCD spectra for the cobalt and iron sites in the two XPt compounds
 and for magnetization along the two high symmetry axis ([001] and 
[100]). The iron and cobalt XMCD show different behavior.
 The picks intensities for the cobalt spectrum lay
 higher for {\bf M}$\parallel$[001] and the
 $L_3$/$L_2$ integrated branching ratio is larger for this axis.
 Contrary, the
 picks intensity for the iron site are larger for {\bf M}$\parallel$[100]
 and the  $L_3$/$L_2$ integrated branching ratio is about constant.}
\end{figure}

Figure \ref{fig5} presents the XMCD spectra for cobalt and iron
atoms for the [100] and [001] magnetization axis. 
We convoluted our theoretical spectra using a Lorentzian
width of 0.9 eV and a Gaussian width of 0.4 eV as proposed by
Ebert,\cite{Ebert} in the case of iron to account for the core hole
effect and the experimental resolution, respectively. The energy
difference between the $L_3$ and $L_2$ peaks is given by the
spin-orbit splitting of the $p_{\frac{1}{2}}$ and
$p_{\frac{3}{2}}$ core states. It is larger in the case of cobalt,
14.8 eV, than for iron, 12.5 eV. The intensities  of the peaks are  comparable
for both atoms, but the cobalt peak-intensities are  larger for the [001]
axis contrary to the iron spectrum. The most important feature of
these spectra is the integrated $L_3$/$L_2$ branching ratio as it
enters the sum rules. It is larger in the case of cobalt site. This is
expected since  the sum rules predict that a larger $L_3$/$L_2$ ratio 
is equivalent to a larger orbital moment. For both atoms the integrated $L_3$/$L_2$
branching ratio is larger for the [001] axis. But the ratio
anisotropy is larger for the cobalt site (1.32 for {\bf M}$\parallel$[001] and
1.17 for {\bf M}$\parallel$[100]) than for the iron site (1.15 for {\bf
M}$\parallel$[001] and 1.14 for {\bf M}$\parallel$[100])
reflecting the larger orbital moment anisotropy of cobalt compared to
iron (see Table \ref{table1}). Especially for iron both orbital moment
and integrated $L_3$/$L_2$ branching ratio are practically the
same for both magnetization axis. As these changes in the XMCD
signal are related to the change in the orbital magnetic moment
between the two magnetization directions. The XMCD anisotropy
should be roughly proportional to the underlying MCA but no relation
exist that connects these two anisotropies. However
they are connected indirectly through the orbital moment 
anisotropy.\cite{bruno} 

\begin{figure}
\begin{minipage}{6.0in}
\epsfxsize=5.0in \epsfysize=4.0in \centerline{\epsfbox{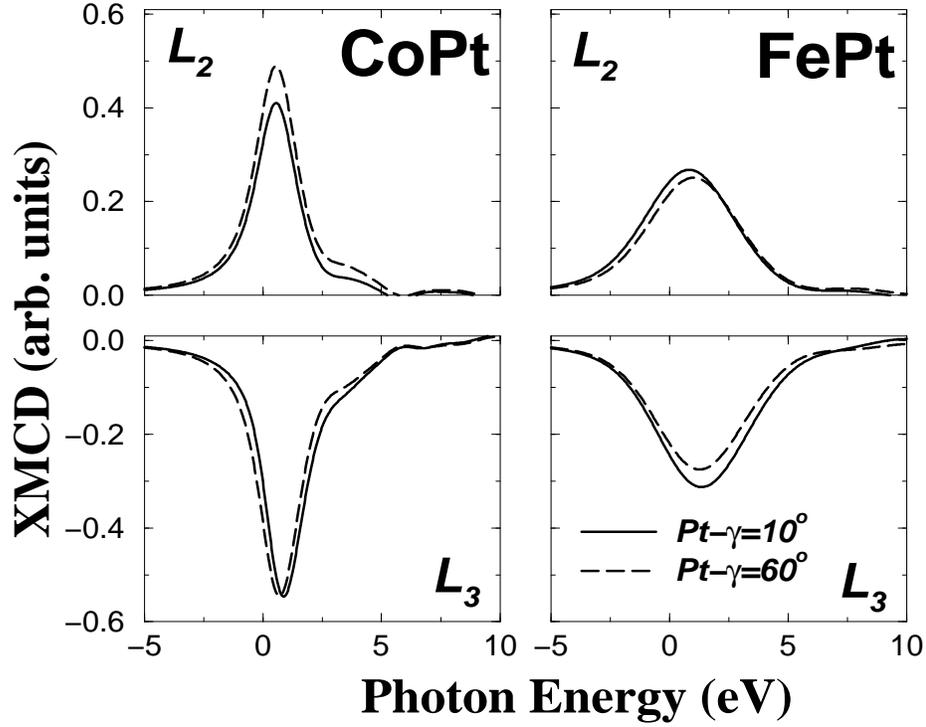}}
\end{minipage}
\caption{ \label{fig9}
XMCD spectra for platinum for two values of the angle $\gamma$ 
between the [001] direction and the spin quantization axis in both
compounds. The differences in the XMCD are important
and the $L_3$/$L_2$ integrated branching ratio changes
considerably with the angle. The intensities for the CoPt compound are
larger reflecting the bigger orbital moments in this compound. The
spin-orbit splitting of the core $p_{\frac{1}{2}}$ and
$p_{\frac{3}{2}}$ states is 1727eV in both compounds. }
\end{figure}

In figure \ref{fig9} we have plotted the platinum XMCD spectra for two
angles $\gamma$=10$^o$ and 60$^o$ for both compounds. The life
time of the core-hole in platinum is smaller than for cobalt so the
broadening used to account for its life time should be larger. We
used both a Lorentzian (1 eV) and a Gaussian (1 eV) to represent
this life time
 and a Gaussian of 1 eV width for the experimental resolution. As in the
case of cobalt and iron the platinum XMCD spectra change with the angle and
depend on the surrounding neighbors. The peak intensities are larger in
CoPt. Also the difference in the intensity due to the anisotropy 
has different sign in the two
compounds. The intensities of  FePt spectra for $\gamma$=10$^o$ are much larger
than for $\gamma$=60$^o$ contrary to the CoPt behavior. As is  the
case for cobalt in CoPt, the platinum site integrated $L_3$/$L_2$ branching ratio
shows larger anisotropy than for platinum in FePt. In CoPt it is 1.49
for $\gamma$=10$^o$ and 1.19 for $\gamma$=60$^o$, while for FePt
it is 1.20 for $\gamma$=10$^o$ and 1.14 for $\gamma$=60$^o$. Here again
the   XMCD follows the anisotropy of the
orbital magnetic moment in these compounds. The energy difference
between the two peaks is 1727 eV for both compounds. The 2$p$
electrons of platinum are deep in energy and  are little influenced
by the local environment, so that their spin-orbit splitting does not
depend on the neighboring atoms of platinum.

\begin{figure}
\begin{minipage}{6.0in}
\epsfxsize=5.0in \epsfysize=4.0in \centerline{\epsfbox{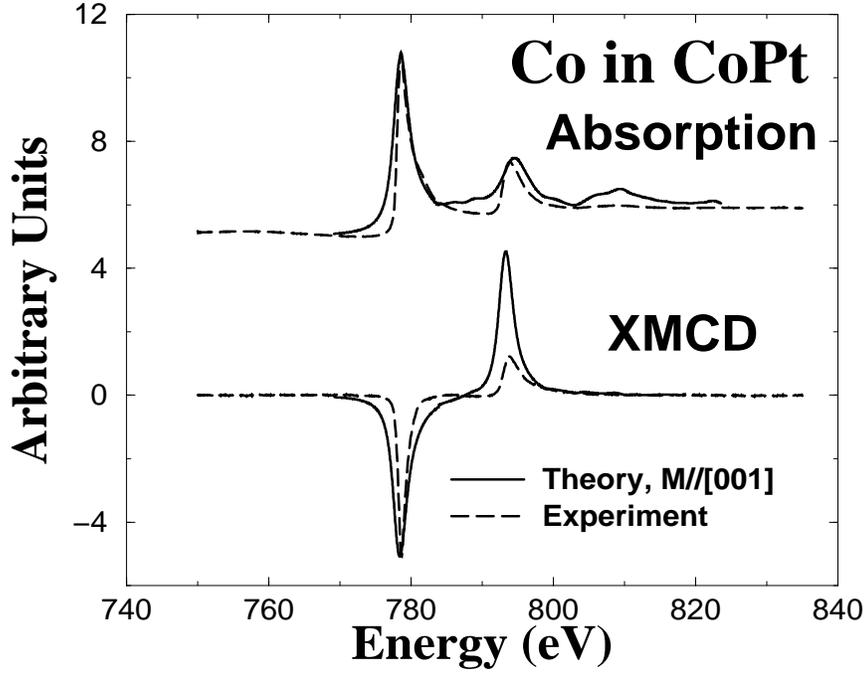}}
\end{minipage}
\caption{ \label{fig6}
Theoretical and experimental absorption and XMCD spectra for the cobalt site 
in CoPt 
for $\gamma$=0$^o$, where $\gamma$ the angle between the [001] 
direction and the spin quantization axis. The theory overestimates the
 absorption at the $L_2$ edge
and  underestimates the $L_3$/$L_2$ integrated branching ratio.
}
\end{figure}

 We expect a better
agreement between the theoretical  and the experimental XMCD spectra 
for the platinum site  than for the cobalt site,
because the core hole is deeper
and would effect less the final states of   the photo-excited
electron. In figure \ref{fig6}, we have plotted the
absorption and the XMCD spectra of cobalt for $\gamma$=0$^o$. We have
scaled our spectra in a way that the experimental and theoretical
$L_3$ peaks in the absorption spectra have the same intensity. The
energy difference between the $L_3$ and  $L_2$ peaks is in good
agreement with experiment. But the intensity of the $L_2$ peak is
larger than the corresponding experimental peak. The high
intensity of the calculated $L_2$ edge makes the theoretical XMCD
integrated $L_3$/$L_2$ branching ratio of 1.32 much smaller  than
the experimental ratio of 1.72. This is because the LSDA
fails to represent the physics of the core hole photo-excited
electron recombination. In the case of 3$d$ ferromagnets the
core hole is shallow and influences the final states seen by 
the photo-excited electron. A formalism that can treat this
electron-hole interaction  has been proposed by Schwitalla and Ebert,\cite{schwit}
but it failed to improve the $L_3$/$L_2$ branching ratio
of XAS of the late transition metals.
Benedict and Shirley\cite{benedict} have
also developed a scheme to treat this phenomena but its
application is limited only to crystalline insulators.

\begin{figure}
\begin{minipage}{6.0in}
\epsfxsize=5.0in \epsfysize=4.0in \centerline{\epsfbox{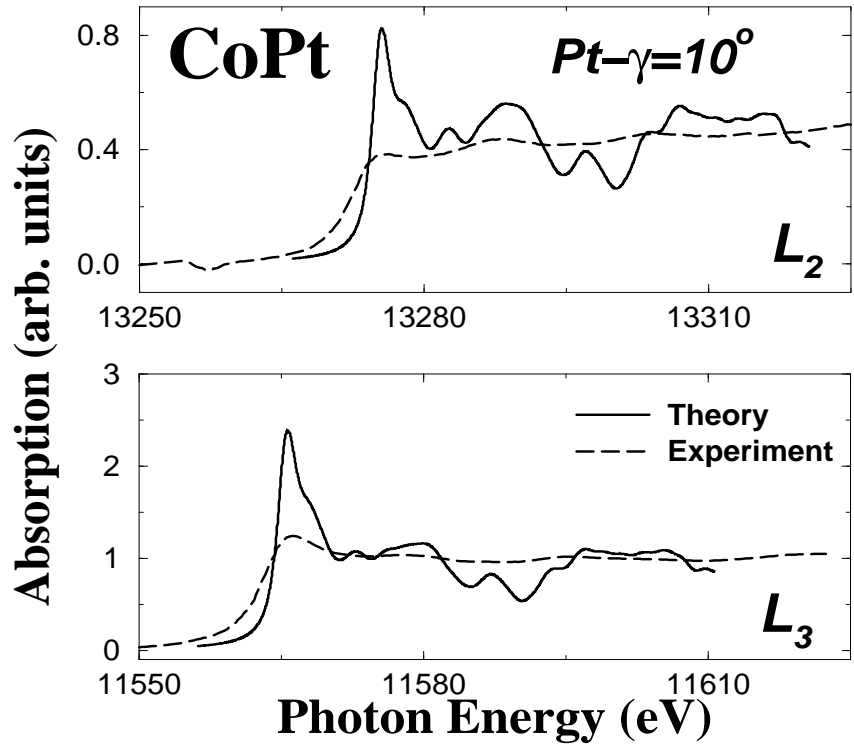}}
\end{minipage}
\caption{ \label{fig7}
Calculated and experimental absorption spectra at the  platinum 
site and for $\gamma$=10$^o$ in
CoPt, where $\gamma$ is the angle between the [001]
direction and the spin quantization axis.
 The theory produces a peak at the threshold of  both
$L_{2,3}$ edges. Both experimental and theoretical spectra have
the same shape for the two edges but $L_3$ edge is much more
intense. Most of the structures in the experimental spectra are washed out 
due to a strong broadening effect.}
\end{figure}

\begin{figure}
\begin{minipage}{6.0in}
\epsfxsize=5.0in \epsfysize=4.0in \centerline{\epsfbox{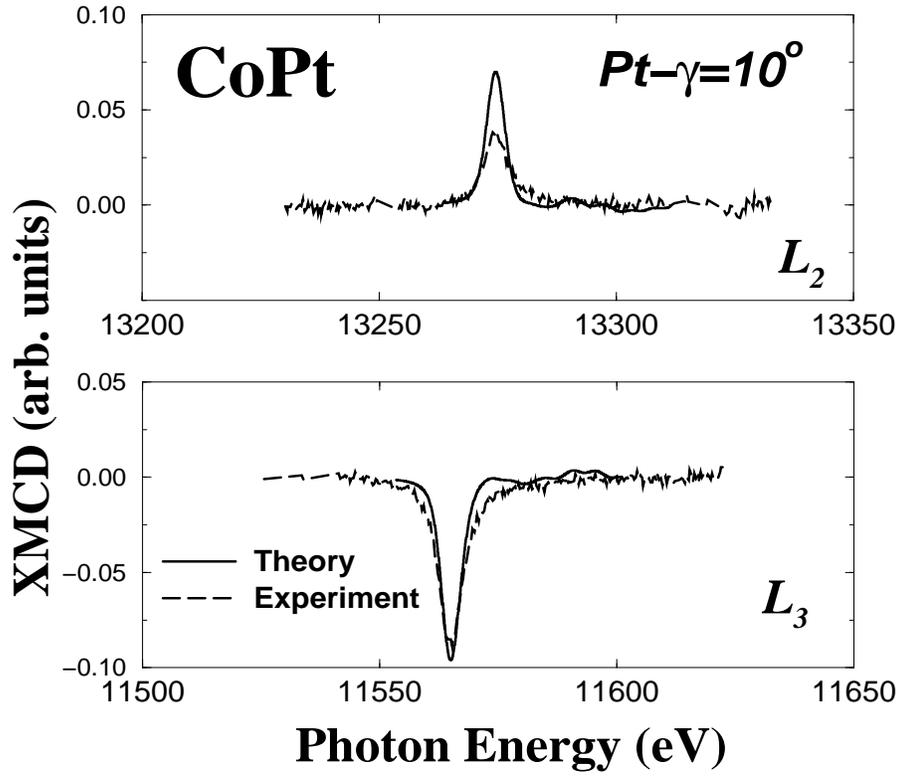}}
\end{minipage}
\caption{ \label{fig8}
Calculated and experimental platinum XMCD spectra at $L_2$ and $L_3$ 
edges for $\gamma$=10$^0$ in
the CoPt compound ($\gamma$ is the angle between the [001]
direction and the spin quantization axis).
 The agreement between the theory and the experiment is
better than in the case of cobalt because the core hole is deeper.
 }
\end{figure}

In figure \ref{fig7} we have plotted experimental and
theoretical total absorptions for the platinum atom in CoPt. For both
$L_2$ and $L_3$ edges the theory gives a sharp peak which does not
exist in experiment. As expected the $L_3$ peak is much more intense
than the $L_2$. In contrast  to what is obtained for cobalt, the
results for the platinum XMCD (see figure \ref{fig8}) show  better
agreement with experiment, due to the fact that the core hole
effect is less intense (core hole much deeper
 compared to cobalt).
 The
experimental and theoretical $L_2$ and $L_3$ edges are separated
by a spin-orbit splitting of the $2p$ core states of 1709 and
1727eV respectively. The width of both $L_2$ and $L_3$ edges is
comparable to  experiment, but the calculated $L_2$ edge is much
larger. This produces a calculated integrated branching ratio of
1.49 which is much smaller than the experimental ratio of 2.66.
Here again the theory is underestimating the branching ratio.

\section{Band Structure and Density of States Anisotropy}
\begin{figure}
\begin{minipage}{6.0in}
\epsfxsize=5.0in \epsfysize=4.0in \centerline{\epsfbox{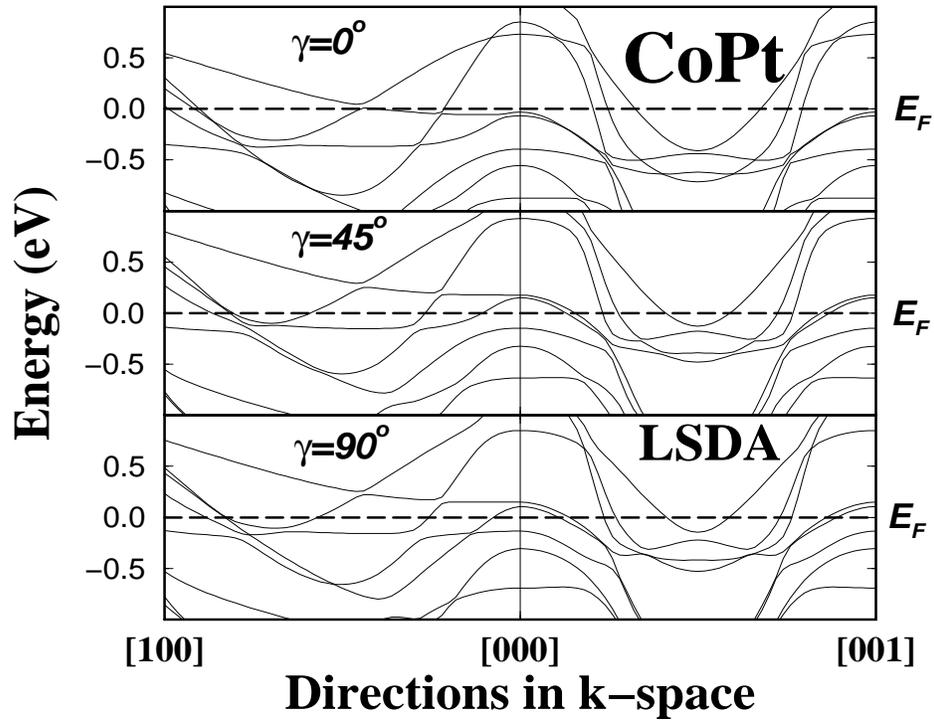}}
\end{minipage}
\caption{ \label{fig10}
CoPt band structure around the Fermi level along two
 high symmetry directions of the Brillouin zone as a function
of the angle $\gamma$ between the [001] direction and the spin quantization axis.}
\end{figure}

\begin{figure}
\begin{minipage}{6.0in}
\epsfxsize=5.0in \epsfysize=4.0in \centerline{\epsfbox{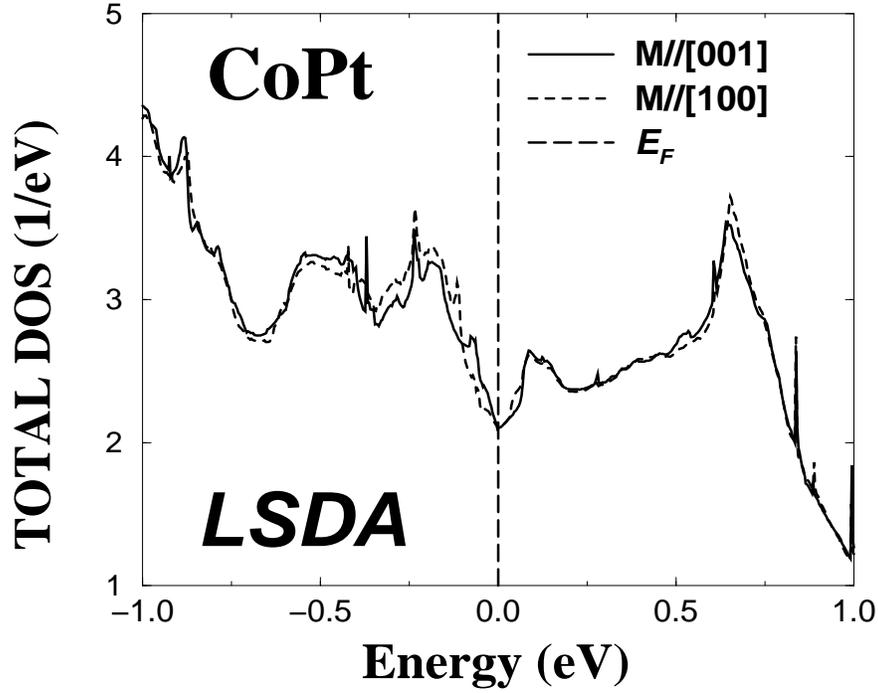}}
\end{minipage}
\caption{ \label{fig11}
Total density of states for CoPt at the vicinity of the 
Fermi level, for the spin quantization axis along the [001] and [100] 
magnetization axis. The states just below the Fermi level suggest that 
the hard axis is favored which led as to conclude that the MCA is due to
the changes in all  the DOS.}
\end{figure}

In figure\ \ref{fig10} we present the band structure along the
[001] and [100] axis in the reciprocal space for different angles
$\gamma$ for the CoPt compound within the LSDA. We know that it  is
essentially the area around the Fermi level that changes 
with respect to the spin quantization axis.
For this reason we have enlarged  a region of $\pm$1 eV
 around the Fermi level. In the first panel we plot the
relativistic  band structure for $\gamma$=0$^o$. In the second and
third panel we have plotted the relativistic band structure for
$\gamma$=45$^o$ and $\gamma$=90$^o$.  We remark
that as the angle increases there are bands that approach the Fermi
level and cross it. However  this information concerns just two
high symmetry directions. 
For this reason   we limited ourselves to the changes in the total DOS. In  figure
\ref{fig11} we notice that just below the Fermi level the 
 DOS for the hard axis is lower than for
the easy axis which seems to favor the 
hard axis. This means that the  anisotropy does not originate from the 
changes ate the vicinity of the Fermi level but from that of the whole
DOS. 
It   is
difficult to investigate this phenomena  by inspection of the changes 
at the vicinity of  the Fermi surface and to 
explain the sign of the MCA . 
Our results confirm the work of Daalderop and
collaborators\cite{daal2} that argued  that not only states in the
vicinity of the Fermi surface contribute to the MAE, as originally
thought,\cite{wang} but states far away make an equally important
contribution. 

\section{Conclusion}

We have performed a theoretical {\em ab-initio} study of the
magnetic properties of the ordered CoPt and FePt fct alloys
systems. The calculated easy axis is the [001] for both compounds
in agreement with other calculations and with experiments on films
which found a strong perpendicular magnetization axis. The density of
states is found to change very little with the direction of the spin 
quantization axis,
 and hence the magnetic moments are isotropic with respect to the
magnetization axis. Contrary to the spin moments, the orbital magnetic
moments decrease with the angle $\gamma$ up to 75$^o$
($\gamma$ is the angle between the  spin quantization
axis and the [001] axis). 

The calculated x-ray magnetic circular dichroism (XMCD)
for all the atoms reflect the behavior of the orbital moments.
Especially platinum resolved spectra present large differences between the
two compounds.  Cobalt XMCD spectra are in agreement with
experiment but as usual the $L_3$/$L_2$ ratio is underestimated by
the theory. The platinum site shows better agreement with experiments
because the core-hole is much deeper than in the case of cobalt.

Finally we showed 
that all the occupied electronic states
contribute to the magneto-crystalline anisotropy and not just
states near the Fermi level for both CoPt and FePt.
\section*{Acknowledgments}
We thank J. M. Wills for providing us with his FPLMTO code. I.G.
is supported by an European Union grant N$^o$ ERBFMXCT96-0089.
The calculations were performed using the SGI Origin-2000
supercomputer of the Universit\'e Louis Pasteur de Strasbourg and
the IBM SP2 supercomputer of the CINES  under grant gem1917.

\begin{table}
\caption{Calculated LSDA and GGA cobalt(iron) orbital magnetic moments
with respect to the angle $\gamma$ between the [001] axis and the spin
quantization axis in the (010) plane. For both atoms the orbital moments
decrease with the angle but values for $\gamma$=90$^o$ do not
follow this trend. Both LSDA and GGA produce  the same trends. The  moments
at the cobalt site 
are larger than at the  iron site. The experimental orbital moments  for the cobalt site
are taken from the Ref. \protect\onlinecite{grange2}. The theory 
underestimates the experimental orbital moments by more than 50\% but
reproduces the correct trends. } \label{table1}
\begin{tabular}{|c|c|c|c|c|c|c|c|}
$\gamma$ & 0$^o$ & 10$^o$ & 30$^o$ & 45$^o$ & 60$^o$ & 75$^o$ & 90$^o$ \\
\hline \hline Fe-LSDA & 0.072 & 0.070 &0.062 &0.051 &0.036 &0.019 &0.070
\\ \hline
Fe-GGA &0.068 &0.067 &0.059 &0.048 &0.034 &0.018 &0.068 \\ \hline \hline
Co-LSDA & 0.109 &0.107& 0.095& 0.076 &0.052& 0.022& 0.061 \\ \hline
Co-GGA &0.088& 0.087 &0.077 &0.063 &0.044 &0.023& 0.061 \\ \hline
Co-Exp & & 0.26 & 0.24 & 0.17 & 0.11 & & 
\end{tabular}
\end{table}

\begin{table}
\caption{Calculated LSDA and GGA  orbital moments of   platinum in FePt and CoPt 
with respect to the angle $\gamma$ between the [001] axis and the spin
quantization axis in the (010) plane.
 For the FePt the GGA produces larger values than LSDA
contrary to CoPt. The moments decrease with the angle but increase sharply 
at  $\gamma$=90$^o$. The moments are significantly larger for
the CoPt compound. Experimental results for platinum in CoPt are taken
from the Ref. \protect\onlinecite{grange2}. The agreement is better
than in the case of cobalt.} \label{table2}
\begin{tabular}{|c|c|c|c|c|c|c|c|}
$\gamma$ & 0$^o$ & 10$^o$ & 30$^o$ & 45$^o$ & 60$^o$ & 75$^o$ & 90$^o$ \\
\hline \hline FePt-LSDA &0.046 &0.045& 0.039& 0.032& 0.023& 0.012 &0.052
\\ \hline
FePt-GGA& 0.049& 0.048 &0.043 &0.035 &0.025 &0.013 &0.058\\ \hline \hline
CoPt-LSDA &0.063& 0.061& 0.054 &0.044& 0.030& 0.014& 0.078 \\ \hline
CoPt-GGA &0.061& 0.060& 0.053 &0.043 &0.031 & 0.016& 0.072 \\ \hline
CoPt-Exp & & 0.09 & & & 0.06 & & 
\end{tabular}
\end{table}

\newpage
%
%
%
\newpage
%
%
%
\newpage
%
%
%
\newpage

%
%
%
\newpage
%
%
%
\newpage
%
%
%

\newpage
%
%
%

\newpage
%
%
%
\newpage
%
%
%

\newpage
%
%
%
\newpage
%
%
%

\end{document}